\documentclass[pteplogo]{ptephy_v1}

\preprintnumber{XXXX-XXXX} %%% %%% Insert preprint number here
\usepackage{hyperref}
\usepackage{amsfonts}
\usepackage{amsmath}
\usepackage{bm}
\usepackage{mathrsfs}
\usepackage{here}
\usepackage{revsymb}
\usepackage{comment}

\newif\ifoneauthor
\oneauthortrue
\newcommand{\unit}[1]{\ {\rm #1}}

\newcommand*{\mpl}{m_{\rm Pl}}

\newcommand{\bmath}[1]{\mbox{\boldmath$#1$}}
\newcommand{\sbmath}[1]{\mbox{\scriptsize\boldmath$#1$}}

\definecolor{mygreen}{RGB}{0,115,0} 
\DeclareMathAlphabet{\mathpzc}{OT1}{pzc}{m}{it}
\usepackage{here} %図を強制的にその位置に出力するためのpackage \begin{figure}[H]でできる。
\definecolor{gray}{gray}{0.4}

\begin{document}

\title{Scalar polarization window in gravitational-wave signals}

%%%% To generate auto affiliation numbers please use \author{}\affil{} command

\author[1*]{Hiroki Takeda}
\affil{Department of Physics, Kyoto University, Kyoto 606-8502, Japan, \email{takeda@tap.scphys.kyoto-u.ac.jp}}
\author[1]{Yusuke Manita}
\author[1]{Hidetoshi Omiya}
\author[1, 2]{Takahiro Tanaka}
\affil{Center for Gravitational Physics and Quantum Information, Yukawa Institute for Theoretical Physics, Kyoto University, Kyoto 606-8502, Japan}

\begin{abstract}
Scalar polarization modes of gravitational waves, which are often introduced in the context of the viable extension of gravity, have been actively searched. However, couplings of the scalar modes to the matter are strongly constrained by the fifth-force experiments. Thus, the amplitude of scalar polarization in the observed gravitational-wave signal must be significantly suppressed compared to that of the tensor modes. Here, we discuss the implications of the experiments in the solar system on the detectability of scalar modes in gravitational waves from compact binary coalescences, taking into account the whole processes from the generation to the observation of gravitational waves. We first claim that the energy carried by the scalar modes at the generation is, at most, that of the tensor modes from the observed phase evolution of the inspiral gravitational waves. Next, we formulate general gravitational-wave propagation and point out that the energy flux hardly changes through propagation as long as the background changes slowly compared to the wavelength of the propagating waves. Finally, we show that the possible magnitude of scalar polarization modes detected by the ground-based gravitational-wave telescopes is already severely constrained by the existing gravity tests in the solar system.
\end{abstract}

\maketitle

% References should be done using the~\cite, \ref, and \label commands

%\setlength{\baselineskip}{10mm}

\section{Introduction}
\label{sec:introduction}
%General relativity (GR) is now widely accepted as the standard theory of gravity through the experimental and observational tests in the weak gravitational fields~\cite{Will1993, Will2005}. However, many alternative theories of gravity have been proposed as extensions of GR motivated from both theoretical and observational aspects~\cite{Will2005, Clifton2012}. Therefore, it is essential to test GR in extreme environments strong gravitational fields or cosmological long scales, where nonlinear and cosmological effects are more significant.

Observations of gravitational waves (GWs) from compact binary coalescences~\cite{Abbott2018, Abbott2020, Abbott2021} by advanced LIGO~\cite{Aasi2015} and advanced Virgo~\cite{Acernese2015} made it possible to test general relativity (GR) in the unprecedentedly extreme regime~\cite{Abbott2019, Abbott2019b, Abbott2020b, Abbott2021b}. 
In alternative theories of gravity, additional polarization modes of GWs are often predicted~\cite{Eardley1973, Will1993}. There exist only two tensor modes (+ and $\times$ modes) in GR, while there exist four additional polarization modes: two vector modes ($x$ and $y$ modes) and two scalar modes ($b$ and $l$ modes), in the theories beyond GR. 
Detection of anomalous modes offers a clear signature of the violation of GR.  
The polarization modes beyond GR from compact binary coalescences have been actively searched in the actual GW data~\cite{Abbott2019, Abbott2019b, Takeda2021, Takeda2022, Hagihara2019, Abbott2020b, Abbott2021b}.

Since the maximum number of polarization modes that can be simultaneously tested is determined by the number of GW detectors, the current search is limited to the case in which only one additional scalar mode is present~\cite{Takeda2018}.  
%The $b$ and $l$ modes are effectively combined by their degeneracy. In addition, given the lack of obvious anisotropic nature in the local gravity of the solar system, it would be reasonable to assume that the GWs do not possess anisotropic features, at least for low-energy excitations. 
Thus, we focus on the scalar polarization modes.
In this paper, we investigate the implication of the experiments in the solar system on the scalar-polarization search in GWs from compact binary coalescences, 
asking the question of whether there exists a window among the extensions of GR 
in which detectable magnitude of scalar modes can be generated. 

First, we look at the generation process. 
The energy loss through the radiation of extra modes modifies the orbital evolution of inspiralling compact binaries and hence the phase evolution of the GW tensor modes~\cite{Chatziioannou2012}. In \cite{Yunes2016}, constraints on the deviation from GR in the energy loss rate are placed by evaluating the phase correction in a model-independent manner. Recently, the possibility of a mixture of polarization modes, {\it i.e.}, the tensor modes along with subdominant scalar modes has been analyzed under the assumption that the waveform of the scalar mode is identical to the inspiral waveform for the tensor polarization modes~\cite{Takeda2022}. In the analysis, the effect of the phase correction due to the scalar-mode radiation is more important than the existence of scalar polarization mode itself for GW170814~\cite{Abbott2017} and GW170817~\cite{Abbott2017b}. 
%{\i.e.}, what is actually constrained is the energy loss due to the scalar radiation. In summary, 
These observational constraints show that the energy flux of the scalar modes from compact binary systems is, at most, that of the tensor modes.

Secondly, we consider the propagation of GWs between the source and the observer.  
We assume that the propagating eigenmodes are described by linear combinations of the six polarization modes. 
Here, our key assumption is that the propagation speed of all propagating modes that we are concerned with is identical to the speed of light. This is because we focus on the possibility that additional scalar modes are detected simultaneously with the tensor modes 
%that have been probed by the polarization tests 
using the GR waveform templates.
Hence, we exclude most of the viable models with an extra scalar field in which screening mechanism~\cite{Khoury2004, Khoury2004b, Vainshtein1972, Deffayet2002, Hinterbichler2010, Hinterbichler2011, Babichev2009, Babichev2013} is crucial to circumvent the experimental constraints. Such models introduce non-vanishing mass or modified kinetic term for the screening to work,  and thus the propagation speed of the extra mode differs from the speed of light. 

Instead of considering concrete examples of modified gravity models, we discuss by introducing a little more generic framework 
which includes polarization mixing, allowing the tensor modes to be transformed to scalar polarization modes during the propagation to avoid the energy-loss constraint from the GW phase evolution. 
We show that the energy flux hardly changes as long as the background changes adiabatically, 
although the amplitude of the polarization may change during GW propagation. 

Finally, we investigate the constraint on the sensitivities of GW detectors on scalar modes from the experiments in the solar system.
The basic idea is that if GW detectors are sensitive to the scalar modes, we should be also able to probe the presence of scalar modes by exploring the additional gravitation force, {\it i.e.}, a fifth force~\cite{Fischbach1998}, which is mediated by scalar modes. The strength of the fifth force has been constrained by various experiments~\cite{Murata2015}. These experiments place a strong upper bound on the coupling between the scalar modes and the matter, implying that the coupling of GW detectors to the scalar modes should be significantly weaker than that to the standard tensor modes. Hence, the contribution of the scalar mode in the GW signal must be suppressed because of the weakness of this coupling.  
%between the scalar modes and the matter.
Combining all, we 
%show that the constraint on the energy loss to the scalar mode is translated to the constraint 
give an upper bound on the possible amplitude of the scalar polarization modes observed by GW detectors.  
%using the constraint on the fifth force by the experiments in the solar system.

This paper is organized as follows.  
In section~\ref{sec:generation}, we consider the GW generation process to evaluate the constraint on the energy loss due to scalar modes from the phase evolution of the inspiral GWs. 
In section~\ref{sec:propagation}, we consider the GW propagation process and formulate general propagating modes, taking into account the possibility of polarization mixing. In section~\ref{sec:observation}, we investigate the GW observation process. 
We discuss two different simplified cases as to the variation of polarization modes and the background field configurations at the position of observers. In the first case, we consider the possibility of observing independent scalar modes 
generated at the source or during propagation. 
%from the source to the detector along with the tensor modes. 
Here, we assume that the local background is isotropic for simplicity. In the second case, we consider the possibility of observing the scalar polarization mode induced by the local anisotropic backgrounds from the tensor polarization modes. Here, we assume that there are only two propagating modes. 
Finally, section~\ref{sec:Discussions-Conclusion} is devoted to the discussions and conclusion. We compare the derived possible amplitude of the scalar mode with the detection limit by the ground-based GW detectors to clarify the detectability of the scalar polarization modes. Throughout this paper, we use the units $c=1$ and $\mpl^2/2:=(16\pi G)^{-1}=1$. Parentheses are used for symmetrization for indices, {\it i.e.}, $A_{(ij)}=(A_{ij}+A_{ji})/2$.

\section{Generation process}
\label{sec:generation}
In this section, we consider the GW generation process and evaluate the constraint on energy loss due to additional scalar modes from the observed phase evolution of the inspiral GWs. 
The modified energy loss from the compact binary system changes the evolution of the binary motion. It results in the phase correction to the inspiral GWs. The modification of the energy flux $\dot{E}$ is often expressed as~\cite{Chatziioannou2012, Yunes2016}
\begin{align}
\dot{E}=\dot{E}_{\rm GR}(1+B_{q} v^{2q})\,,
\label{eq:Bq}
\end{align}
where $\dot{E}_{\rm GR}$ is the energy flux in GR, $v$ is the binary orbital velocity, and $B_q$ is the phenomenological deviation parameter at $q$-th post-Newtonian (PN) order for the energy loss from the compact binary system. 
Through the stationary phase approximation, we can calculate the phase correction by extra energy loss due to radiation of the $\ell$-th harmonic of the inspiral GW~\cite{Chatziioannou2012},
\begin{align}
\delta\Psi^{(\ell)}_q=\frac{15}{64}B_{q}\frac{\ell}{(4-q)(5-2q)}\eta^{-2q/5}(2\pi\mathcal{M}F)^{(2q-5)/3}\,,
\label{eq:phase_correction}
\end{align}
where $\mathcal{M}$ is the chirp mass and $F$ is the orbital frequency that is related to the GW frequency as $f=\ell F$.

The phase corrections of inspiral GWs have been constrained in terms of the deviation parameters from the phase coefficients in GR introduced at each PN order as $(1+\delta\hat{p}_{q})p_{{\rm GR}, q}$, where $p_{{\rm GR}, q}$ is the coefficient in GR and $\delta\hat{p}_{q}$ is the deviation parameter at the $q$-th PN order~\cite{Arun2006, Arun2006b, Mishra2010, Yunes2009, Li2012, Cornish2011, Meidam2018}. The constraints on $\delta\hat{p}_q$ can be translated into the constraint on $B_{q}$ through Eq.~\eqref{eq:phase_correction}.  
For example, the constraint on $B_{-1}$ corresponding to the dipole radiation and on $B_{0}$ corresponding to the quadrupole radiation were reported for the two binary black hole events as~\cite{Yunes2016}
\begin{align}\label{eq:constraintB}
	B_{0}\lesssim &10^{-1}~, & B_{-1}\lesssim & 10^{-2}~.
\end{align}
 
Since vector or scalar modes are not allowed in GR, such additional polarization modes contribute to $B_q$,  if they exist. Thus, the constraints~\eqref{eq:constraintB} place upper bounds on the energy flux of additional polarization modes. 
However, since how the polarization modes are normalized is unspecified yet, the constraint~\eqref{eq:constraintB} does not directly place an upper bound on the amplitude of additional polarization modes, 
which are explored in the polarization tests. 
The observed GWs emitted from the inspiralling binaries did not show any signature of a large amount of energy leakage to extra modes.
The possible energy carried by additional modes is, at most, that of the tensor modes as far as the observed phase evolution matches the GR prediction in terms of $\delta\hat{p}_q$. 

\section{Propagation process}
\label{sec:propagation}
In this section, we formulate the GW propagation process with general mixing of polarization modes. 
We show that the energy flux does not change in the slowly-varying adiabatic background, although the amplitude of each polarization mode can change during GW propagation.

In general, any independently propagating modes in $\hat{e}_z$ direction would be locally
given by the linear combination of fundamental polarization modes, 
\begin{align}
    h^I_{ij}
    &= \sum_{A} h^I_A e^{A}_{ij}~,
\end{align}
where $I$ is the label for the canonically normalized propagating modes and $A \in\{+,\times, x, y, b, l\}$ is the label for the polarization modes. Lower case latin indices run over the spatial directions.  
In the synchronous gauge $h_{0\mu}=0$, polarization basis set $\{e^A_{ij}\}$ is given by~\cite{Takeda2018}
\begin{align}
    e^{+}_{ij} &= \hat{e}_{x,i}\hat{e}_{x,j}-\hat{e}_{y,i}\hat{e}_{y,j}\,, & e^{\times}_{ij} &= \hat{e}_{x,i}\hat{e}_{y,j}+\hat{e}_{y,i}\hat{e}_{x,j}\,,
    \label{eq:tensormodes}
    \\
    e^{x}_{ij} &= \hat{e}_{x,i}\hat{e}_{z,j}+\hat{e}_{z,i}\hat{e}_{x,j}\,, & e^{y}_{ij} &= \hat{e}_{y,i}\hat{e}_{z,j}+\hat{e}_{z,i}\hat{e}_{y,j}\,,
    \\
    e^{b}_{ij} &= \hat{e}_{x,i}\hat{e}_{x,j}+\hat{e}_{y,i}\hat{e}_{y,j}\,, & e^{l}_{ij} &= \sqrt{2}~\hat{e}_{z,i}\hat{e}_{z,j}\,.
    \label{eq:scalarmodes}
\end{align}
Here, $\{\hat{e}_{x,i}, \hat{e}_{y,i}, \hat{e}_{z,i}\}$ are the mutually orthogonal unit basis vectors. We refer to $+$ (plus) and $\times$ (cross) modes as tensor modes, $x$ (vector x) and $y$ (vector y) modes as vector modes, and $b$ (breathing) and $l$ (longitudinal) modes as scalar modes. 
%Note that the polarization modes are observable and defined as the gauge invariant quantities $h_P$ through irreducible decomposition of tensor representation for the metric perturbation under SO(2) transformation around a fixed axis pointing the GW propagation direction~\cite{Poisson2014}. 
It is straightforward to generalize the propagation direction in the above expressions~\eqref{eq:tensormodes}-\eqref{eq:scalarmodes}. 
Now, we describe the metric perturbation as 
\begin{equation}
 h_{ij}=\sum_I \phi_I h^I_{ij}+\mbox{(c.c.)}\,,
 \label{eq:basicexpansion}
 \end{equation}
where (c.c.) represents the complex conjugate. In the following, assuming slowly varying backgrounds and modes $\phi_I$ propagating at the speed of light. 
%we show that the quadratic action of the relevant modes 
%can be approximated by the standard form 
%\begin{equation}
% S=\int d^4x \frac{\sqrt{-g}}{2}\sum_I \left( -g^{\mu\nu}\partial_\mu \phi^{*I} \partial_\nu \phi^I -\tilde C_{I\! J}(\omega)\phi^{*I}\phi^J\right)\,, 
% \label{eq:standardform}
%\end{equation}
%along the propagation path, where $\omega$ is the rotation velocity of the phase. 

%We shall provide a justification for assuming this form of quadratic action. 
When we have $n$ real degrees of freedom, the quadratic action would be generally written as 
\begin{align}\label{eq:generic2nd}
    S^{(2)}=\int  d^4x \frac{\sqrt{-g}}{2} \left( -A^{\mu\nu}_{I\!J}(x) \partial_\mu \phi^{I*} \partial_\nu \phi^J 
    -B_{I\!J}^\mu(x) \left\{(\partial_\mu\phi^{I*})\phi^J-\phi^{I*}(\partial_\mu\phi^J)\right\} -C_{I\!J}(x) \phi^{I*}\phi^J\right)\,, 
\end{align}
where we neglect the terms $\propto \phi^I\phi^J, \phi^{I*}\phi^{J*}$ that rapidly oscillate when we substitute $\phi^I=\hat\phi^I e^{-ik_\mu x^\mu}$ assuming that $\hat\phi^I$s are slowly varying functions.
From the condition that the action should be real, 
we find that $A^{\mu\nu}_{I\!J}$ is a symmetric spacetime tensor and $C_{I\!J}$ is a spacetime scalar, which are Hermitian matrices with respect to $I,J$-indices, while 
$B^\mu_{I\!J}$ is a spacetime vector and an anti-Hermitian matrix with respect to $I,J$-indices. The Hermitian part $\tilde B^\mu_{I\!J} \left\{(\partial_\mu\phi^{I*})\phi^J+\phi^{I*}(\partial_\mu\phi^J)\right\}$ can be 
absorbed by the $C_{I\!J}$-term using integration by parts. 
We decompose $A^{\mu\nu}_{I\!J}$ into the part 
proportional to $g^{\mu\nu}$ and the remainder as 
\begin{align}
 A^{\mu\nu}_{I\!J}=A^{0}_{I\!J}g^{\mu\nu}+\delta\! A^{\mu\nu}_{I\!J}\,. 
\end{align}
Then, under the change of variables
\begin{equation}
    \phi^I\to \phi'{}^I=(\Lambda^{-1})^I_{~J} \phi^J\,, 
\end{equation}
the coefficients in the action transform as 
\begin{align}
\label{eq:transformation_A_B_C}
  A^{\mu\nu}_{I\!J}\to A'{}^{\mu\nu}_{I\!J}=&A^{\mu\nu}_{KL}\Lambda^{\!*K}_{~~I}\Lambda^{L}_{~J}\,,\cr
  B^\mu_{I\!J}\to B'{}^\mu_{I\!J}=&\left[B^\mu_{KL}+
  \frac{A^0_{KM} \nabla^\mu \lambda^{M}_{~L}+\delta\! A^{\mu\nu}_{KM} \nabla_\nu \lambda^{M}_{~L}-\mbox{(h.c.)}}{2}\right]\Lambda^{\!*K}_{~~I}\Lambda^{L}_{~J}\,,\\
    C_{I\!J}\to C'_{I\!J}=&C_{KL}\Lambda^{\!*K}_{~~I}\Lambda^{L}_{~J}+\left[B^\mu_{KL}(\nabla_{\mu}\Lambda^{\!K*}_{~~I})\Lambda^{L}_{~J}+\mbox{(h.c.)}\right]\cr
    &\quad+\frac{1}{2} \left[ A^{\mu\nu}_{KL}(\nabla_\mu\Lambda^{\!K*}_{~~I})\nabla_\nu\Lambda^{L}_{~J}+\mbox{(h.c.)} \right]\,,
\end{align}
where (h.c.) represents the Hermitian conjugate, and 
\begin{equation}
    \nabla^\mu \lambda^{M}_{~L}:=(\nabla^\mu \Lambda^{M}_{~J})(\Lambda^{-1})^{J}_{~L}\,.
\end{equation}$\nabla^{\mu}\lambda^{M}_{~L}$ is a generic complex matrix when $\Lambda^{M}_{~L}$ is a generic complex matrix. 
 
%Below, we show that the generic quadratic action \eqref{eq:generic2nd} can be transformed 
%into the standard form \eqref{eq:standardform}, 
First, we transform the Hermitian matrix $A^0_{I\!J}$ to the identity matrix $I_{I\!J}$ by choosing the matrix $\Lambda^{I}_{~J}$. 
%After that, 
%we can set $k_\mu B^\mu_{I\!J}={\cal O}(\delta\! A^{\mu\nu}_{KM} \nabla^\nu \lambda^{M}_{~L})$ along the propagation path where $k_\mu$ is the wave number of the propagating wave, by applying an additional unitary transformation specified by  
%\begin{align}
%\Lambda^{I}_{~J}=\mbox{P Exp}\left[-\frac{k_{\mu}}{\omega}\int^{u}_{u_0} B^{\mu}_{I\!J}du' \right]\,,
%\end{align}
%Finally, %we consider to diagonalize $C_{I\!J}$. 
%Roughly speaking, 
%this additional transformation would break the condition $k_{\mu} B^\mu_{I\!J}=0$, \smc{but the violation should be proportional to the derivative of $C_{I\!J}$, which would be relatively suppressed compared to the contribution coming from the remaining $C_{I\!J}$ term.} 
%Then, each real eigenvalue of a Hermitian matrix $C_{I\!J}$ gives the mass squared of each mode, which must be negligibly small, as long as the propagation speed is sufficiently close to the speed of light.
%Hence, 
%we show that we can safely assume that $C_{I\!J}=0$. 
Then, the equation of motion reduces to
\begin{align}
    \nabla^\mu\nabla_\mu\phi^I-\tilde C^I_{~J}\phi^J=0\,,  
    \label{eq:eom_c}
\end{align}
where 
\begin{align}
    \tilde C_{I\! J}= -k_\mu k_\nu \delta\! A^{\mu\nu}_{I\! J}
    +2i k_\mu B^{\mu}_{I\! J}+C_{I\! J}\,.
\end{align}
Here, we adopt the form satisfying the conditions $A^0_{I\!J}=I_{I\!J}$ and drop the terms that become higher order in $1/\omega L$, where $\omega$ is the angular velocity of the phase, and $L$ is the coherence length of the spatial variation of $\tilde C_{I\!J}$. Namely, we neglect the terms in which the derivative operators act on $\hat\phi^I$, except for the term coming from $A^0_{I\!J}$. 
This can be justified if 
the background quantities $\delta\! A^{\mu\nu}_{I\!J}, B^\mu_{I\!J}$ and $C_{I\!J}$ change slowly compared to the wavelength of the propagating waves.
We note that $\tilde C_{I\!J}$ is a Hermitian matrix.  

Next, we solve the equation of motion perturbatively along the path of the GW. Let us choose the null coordinates $u$ and $v$ on the homogeneous and isotropic universe where $u$ is the normalized coordinate along the path that satisfies $dx^{\mu}/du=k^{\mu}/\omega$ and $v$ is the null coordinate pointing the direction paired to $u$.  We assume the form of the solution
\begin{align}
    \phi^I=\hat\phi^I e^{-ik_\mu x^\mu}\,,
\end{align}
where $\hat\phi^I$s are slowly varying functions given by
\begin{align}
\hat\phi^I=\tilde{U}^{I}_{~J}\hat \phi_0^J\,.
\label{eq:unitaryrotation}
\end{align}
%$\phi'^{I}:=\phi_0^I e^{ik_\mu x^\mu}$ is the solution for the equation of motion without $\tilde C_{I\!J}$-term and 
% $\hat{\phi}_{0}^{J}$ is an initial vector at the source position. 
Here, $\hat{\phi}_0^{J}$ is an initial vector at the source position and $\tilde{U}^{I}_{~J}$ is a certain matrix on the field space. We expand the transformation matrix $\tilde{U}_{I\!J}$ as
\begin{align}
    \tilde{U}^{I}_{~J}={U}^{I}_{~J}+{U}^{I}_{~K}{V}^{K}_{~~J}\,.
    \label{eq:tilde_U_for_C}
\end{align}
By assuming that $\tilde{C}^{I}_{~J}$ is small, we obtain the equation of motion for the leading order correction ${U}^{I}_{~J}$ as
\begin{align}\label{eq:Uleading}
2i\omega\partial_u U=\tilde{C} U\,,
\end{align}
where we adopt the matrix notation, for brevity.
The solution to Eq.~\eqref{eq:Uleading} is formally given by
\begin{align}
{U}:=\mbox{P Exp}\left[\int^{u}_{u_0} \frac{\tilde C}{2i\omega}du' \right]\,.
\end{align}
Here, ``P'' represents the path ordered product and $u_0$ is the initial value of $u$ corresponding to the source position. 
%Thus, $B^\mu_{I\!J}$-term is eliminated. %does not affect the propagation of the waves.  %Here, P stands for the usual path-ordered product again. 
If we neglect $V^K_{~~J}$ in the second term in Eq.~\eqref{eq:tilde_U_for_C}, 
$\tilde U^I_{~J}(u)$ becomes a path ordered product of exponentiated anti-Hermitian matrices and hence a unitary matrix. Therefore, the norm of the 
vector $\hat \phi^I$ is conserved at this level of approximation. 
%and $A_{I\!J}=I_{I\!J}$ still holds. 

Let us evaluate the phase shift caused by $U^I_{~J}$, {\it i.e.}, 
$\varphi:=\left\vert\mbox{Tr}\left[\mbox{Log}(U^I_{~J})\right]\right\vert$.
Here, we assume that the phase shift is given by the summation of the contributions from mutually independent domains whose number is of $O(d_L/L)$. 
Then, the magnitude of the phase shift can be estimated as 
$\varphi\sim\bigl\vert\sqrt{L d_L} \tilde C^I_{~I}/\omega\bigr\vert$, where $d_L$ is the comoving distance from the source to the detector. 
The phase shifts that depend on the frequency as $\propto \omega$ and $\omega^{-1}$ modify the GW arrival time and 
the GW waveform, respectively. 
In the same manner as we give a constraint on the graviton mass using the GW observational data~\cite{Abbott2021b},
the amplitude of these phase shifts is also observationally constrained as $\varphi\lesssim 1$. 
If this condition is largely violated, the detectability of extra polarization modes simultaneously based on the templates in GR would be significantly reduced. Since $\hat\phi^J_0$ does not have a specific form in general, the condition $\varphi\lesssim 1$ 
implies that the unitary matrix $U^I_{~J}$ should be sufficiently close to $I^I_{~J}$ even after the propagation over cosmological distances. 
Namely, all the rotation angles of the unitary matrix $U^I_{~J}$ are less than $O(1)$. 
The only exceptions are the contributions from $\delta\! A^{\mu\nu}_{I\! J}$ in case $\delta\! A^{\mu\nu}_{I\! J}\propto I_{I\! J}$ and $B^\mu_{I\!J}$, which are proportional to $\omega^2$ and $\omega$ in $\tilde C_{I\!J}$, respectively.
The arrival times are shifted in the same way for all modes by the part of $\delta\! A^{\mu\nu}_{I\! J}$ that is $\propto I_{I\! J}$, 
which describes nothing but the standard gravitational lensing effect. 
The contribution from $B^\mu_{I\!J}$ modifies the phase in a frequency independent manner. 
Therefore, even if $\varphi$ originating from $B^\mu_{I\!J}$ is large, we can still detect the signal using GR templates. 
%Therefore, the violation of $k_{\mu}B^{\mu}_{I\!J}=0$ would be sufficiently small as far as $\omega\ll d_L$.} 
%On the other hand, in the case (ii) there is no significant modification of the waveform, if the initial waveforms are common among different modes except for the constant phase shift. Therefore, the magnitude of $C^I_{~J}$ is not strongly constrained. 

%the propagation speed of all $n$ degrees of freedom is almost identical to the speed of light and 
Next, we consider the next to leading order correction, $V^{I}_{~J}$.
Substituting the ansatz \eqref{eq:tilde_U_for_C} into the equation of motion, 
the equation to solve is reduced to
\begin{align}
\nabla^{\mu}\nabla_{\mu} U+2i\omega  U(\partial_{u}V) =0\,, 
%+(\nabla^{\mu}\nabla_{\mu}U)V+2(\nabla^{\mu}U)(\nabla_{\mu}V) +U(\nabla^{\mu}\nabla_{\mu}V) 
\end{align}
where we adopt the matrix notation, for brevity. $\nabla^\mu\nabla_\mu$ is 
decomposed into $\partial_u \partial_v$ and ${}^{(2)}\!\triangle$, which represents the Laplacian in the two spatial dimensions perpendicular to the propagation direction. 
We count the magnitude of each term assuming that $\partial_u \sim 1/d_L$ and 
the differentiation in the other directions $\sim 1/L$, when they act on $U$ or $V$. 
%Here, $L$ is the typical length scale of variation of $C$. 
Assuming $L/d_L\ll 1$, %and $\varphi\lesssim 1, 
we find that the leading order of $V$ is determined by solving
%\begin{align}
%\left[\nabla^{\mu}\nabla_{\mu} U+2i\omega  U(\partial_{u}V) \right]\simeq0\,.
%\end{align}}
\begin{align}
2i\omega (\partial_{u}V) \simeq - U^{-1} {}^{(2)}\!\triangle U\,.
\end{align}
%where \smc{we neglected the subdominant terms coming from the first term in Eq.~\eqref{eq:eom_c}} and 
%expressed the d'Alembertian in the four-dimensional space-time in the light-cone coordinate. 
The equation can be solved formally as
\begin{align}
V\simeq & -\frac{1}{2i\omega}  \left[ 
%\smc{\int^{u}_{u_0} du' U^{-1}(u')(\partial_v W)(u') U(u') +\int^{u}_{u_0} du' U^{-1}(u')W(u')(\partial_vU)(u')}  \right. \cr
%&  
\int^{u}_{u_0} du' \int^{u'}_{u_0} du'' U^{-1}(u') U(u')U^{-1}(u'')\frac{{}^{(2)}\!\triangle \tilde C(u'')}{2i\omega} U(u'')U^{-1}(u_0)\right. \cr
& \qquad\qquad +2 \int^{u}_{u_0} du'\int^{u'}_{u_0} du'' \int^{u''}_{u_0} du''' U^{-1}(u')U(u')U^{-1}(u'') \frac{{}^{(2)}\!\nabla^{i} \tilde C(u'')}{2i\omega} U(u'')\cr
&\qquad\qquad\qquad \left. \times U^{-1}(u''') \frac{{}^{(2)}\!\nabla_{i} \tilde C(u''')}{2i\omega} U(u''')U^{-1}(u_0)  \right] \cr
\simeq &\frac{1}{4\omega^2} \left[%\int^{u}_{u_0} du' \partial_v \tilde C(u') \right. - 
\int^{u}_{u_0} du' (u-u')  {}^{(2)}\!\triangle {\cal C}(u')  -\frac{i}{\omega} \int^{u}_{u_0} du' (u-u')  \int^{u'}_{u_0} du''
{}^{(2)} \nabla^{\mu} {\cal C}(u'){}^{(2)} \nabla_{\mu} {\cal C}(u'')  \right]\,, \cr
\label{eq:V}
\end{align}
where we introduced 
${\cal C}:=U^{-1} \tilde C U$, which is analogous with the interaction picture field in the quantum field theory. 
%Now, we evaluate the magnitude of $V$ %in Eq.~\eqref{eq:tilde_U_for_C} 
%This contribution includes the lensing effects. 
%As is expected already before the integration, %
The magnitudes of the two terms in the last line of Eq.~\eqref{eq:V} would be estimated as 
$O(d_L\varphi/\omega L^2)$ and $O(d_L\varphi^2/\omega L^2)$, 
respectively. 

First, we consider the case in which $\varphi\lesssim 1$ is satisfied. 
In this case, we find that significant amplitude modification or phase shift due to $V$ 
can occur only when $d_L/\omega L^2\gg 1$. 
Namely, $V$ can be neglected as long as   
\begin{equation}
   L\gtrsim 0.3\mbox{pc}\left(\frac{\omega/2\pi}{30\mbox{Hz}}\right)^{-1/2}
        \left(\frac{d_L}{3\mbox{Gpc}}\right)^{1/2}\,. 
        \label{eq:scalelimit}
\end{equation}
If the contribution of $V$ can be neglected, the effect of ${\cal C}$ can 
be absorbed by the unitary rotation given by \eqref{eq:unitaryrotation}. 
Hence, we can safely assume the canonically normalized form of the quadratic action 
\begin{equation}
 S=\int d^4x \frac{\sqrt{-g}}{2}\sum_I \left( -g^{\mu\nu}\partial_\mu \phi^{I*} \partial_\nu \phi^I \right)\,, 
\label{eq:standardform}
\end{equation}
along the propagation path. 

When the condition \eqref{eq:scalelimit} is violated, we need to consider the wave optics to correctly take into account the effect of $V$. The effective size of extension of the observed waves during propagation $\ell$ would be estimated by the condition that the difference of the path length becomes $O(1)$. This means that $\ell\sim \sqrt{d_L/\omega}$. Therefore, when $d_L/\omega L^2\gg 1$, $L$ becomes much smaller than $\ell$ and the background is rapidly changing compared to the effective size of the beam width. In such situations, smearing over the effective size $\ell$ will eliminate the components in $\tilde C$ shorter than the length scale $\ell$. Hence, even in this case any significant modification due to $V$ would be unexpected.
%

%of Eq.~\eqref{eq:standardform}, as far as 
%the condition~\eqref{eq:scalelimit} holds. 
When $\varphi\gg 1$, which can be realized in the case of the $\delta\! A^{\mu\nu}_{I\! J}\propto I_{I\! J}$-term dominance or $B$-term dominance, amplification can happen. The amplitude change is related to the non-unitary nature of $\tilde U$. In the present perturbative analysis, this amplification is caused by the Hermitian component of $V$, which is identified as 
\begin{align}
\frac{V+V^\dag}{2}
\simeq \frac{1}{4\omega^2}  &\left(\int^{u}_{u_0}(u-u') du' {}^{(2)}\!\triangle {\cal C}(u')\right.\cr  
&\left.-\frac{i}{\omega} \int^{u}_{u_0} (u-u') du' \int^{u'}_{u_0} du''
\left[{}^{(2)} \nabla^{\mu} {\cal C}(u'), {}^{(2)}\nabla_{\mu} {\cal C}(u'')\right]  \right)\,. 
\label{eq:V2}
\end{align}
The amplification due to $\delta\! A^{\mu\nu}_{I\! J}\propto I_{I\! J}$-term is the result of the standard gravitational lensing amplifying all modes in the same way. 
%The amplification due to $B$-term is a phenomena similar to the gravitational lensing. 
On the other hand, since ${\cal C}\propto \omega$ in the case of the $B$-term dominance, 
the magnification factor due to $B$-term is frequency dependent, {\it i.e.}, $\propto 1/\omega$. At the same time, the frequency dependent phase shift also occurs owing to the contribution from the anti-commutator, 
\begin{align}
\frac{V-V^\dag}{2}
\simeq &-\frac{i}{4\omega^3} 
\int^{u}_{u_0} (u-u') du' \int^{u'}_{u_0} du''
\left\{{}^{(2)} \nabla^{\mu} {\cal C}(u'), 
{}^{(2)}\nabla_{\mu} {\cal C}(u'') \right\}\,,  
\end{align} 
which leads to the modification of the waveform. 
Hence, the discussion above for the phase shift caused by $U^I_{~J}$ applies and the $B$-term should be small enough to be consistent with the observed GW phase.
When $\delta\! A^{\mu\nu}_{I\! J}\propto I_{I\! J}$-term is 
large but $B$-term is also present, we may need to worry about the cross term 
between these two. 
However, while the anti-commutator of these two can lead to the frequency independent phase shift, but the commutator related to the amplification vanishes. 
Thus, the cross term does not contribute to the amplification. 
Therefore, 
it seems difficult to have a large magnification of specific modes without changing the waveform. 

To conclude, although the amplitude of $h_{ij}$ may change during propagation, as long as the change of the background is adiabatic,  
the energy flux carried by the propagating waves cannot be altered without waveform or arrival time deformation, except for the effect of the usual redshift and gravitational lensing. Here, adiabatic means that the temporal and spatial scales of the gradient of the background fields are much longer than the wavelength of GWs.
We should note that there is no change in the energy flux other than the effect due to the redshift, 
even for the case of well-known examples of the amplitude variation during propagation, such as the Chern-Simons gravity~\cite{Dyda:2012rj} and a scalar field with non-canonical 
kinetic term. %in which the apparent magnitude of the wave changes.
This magnification or de-magnification is due to the difference between the forms of $h^I_{ij}$ 
at the source and at the observer. 

The above argument does not apply if the background is not smooth enough compared with the wavelength of the propagating modes, 
but such a situation seems to be quite unlikely under the condition that the propagation speed is very close to the speed of light. 
As one exceptional possibility, a rapid change of the background might be realized by the presence of a domain wall network in the universe~\cite{Vilenkin2000}, which we do not pursue here.
%However, to make the wall thin enough, the wall tension should be large, 
%which leads to the problem of the dominance of the energy density of the domain wall in the universe.
Before closing this section, 
we would like to stress that we have established the conservation of the flux during the propagation, 
or equivalently the conservation of the amplitude of the canonically normalized field $\phi_I$. 
We only use this fact in the succeeding discussion. 

\section{Observation process}
\label{sec:observation}
In this section, we discuss the detectability of the scalar modes under two different settings regarding the number of polarization modes and the presence of an anisotropic background field. 
First, we consider the case in which an additional scalar mode propagates from the source to the detector along with the tensor modes. Here, we assume that the local background at the position of the observer is isotropic. Secondly, we consider the case in which we observe the apparent scalar mode induced by the anisotropies of the local background field, although there are only two propagating modes emitted from the GW source.
In principle, we can consider the case in which the two mechanisms are simultaneously at work. However, since there does not seem to be any synergistic effect from the coexistence of the two mechanisms, the constraints on the magnitude of the scalar polarization would not be altered.

\subsection{Detectability of independently propagating scalar modes}
Here, we consider the possibility of detecting an additional scalar polarization mode propagating independently of the tensor modes at the location of observers. 
We assume that there are no background anisotropies in the solar system in the gravitational sector. 
In addition, we assume that there is only one additional massless propagating mode with scalar polarization components.
Thus, we would be able to safely assume 
that $h_{ij}$ can be expanded as 
\begin{equation}
    h_{ij}=\sum_{I=1, 2, 3} \psi_I h^I_{ij}\,,
\end{equation}
where each polarization basis is given by
\begin{align}
     & h^{1}_{ij}=e^+_{ij}\,,\qquad
    h^{2}_{ij}=e^\times_{ij}\,,\cr
    & h^{3}_{ij}={\zeta}_b e^b_{ij}+{\zeta}_l e^l_{ij}\,,
    \label{eq:observerbasis}
\end{align}
with ${\zeta}_b$ and ${\zeta}_l$ are some complex constants, which can depend on 
$k=|\vec k|$. 
In order to separate the observation process from the propagation process, we express the propagating modes during propagation as $\phi_I$ while those in the observation process or in the solar system as $\psi_I$.
The new variables $\psi_I$ take the same standard form of the action 
\eqref{eq:standardform} as $\phi^I$ for small perturbation. 
If we have the other scalar degrees of freedom, we can generalize the model only by extending $ h^3_{ij}$, ${\zeta}_b$, and ${\zeta}_l$, to $h^{I}_{ij}$, ${\zeta}^I_b$, and ${\zeta}^I_l$, respectively. 
The extension simply gives additional deviations from GR in the metric perturbation to make the model more inconsistent with the observations. 
There might be other massless propagating degrees of freedom, but we neglect their possible existence here. 
Including extra degrees of freedom will not change the following discussion basically. 

In order to maintain the weak equivalence principle, 
we assume the universal coupling between the metric perturbation and the effective matter stress energy tensor as 
\begin{equation}
    S_{\rm int}=\int d^4x \frac{\sqrt{-g}}{2} h_{\mu\nu}T^{\mu\nu}\,. 
    \label{eq:universal_coupling}
\end{equation}
Here, we assume the existence of the effective stress energy tensor that satisfies the conservation law with respect to the background Minkowski metric $T^{\mu\nu}_{~~,\nu}=0$, 
which means that $T^{\mu\nu}$ includes the contribution from the second-order gravitational perturbation. 
Then, we can calculate the metric perturbation in the synchronous gauge induced by a given stress energy tensor $T_{\mu\nu}$ as
\begin{align}
    \tilde{h}_{ij}(k)=\tilde{G}_{ijkl}\tilde{T}^{kl}(k)\,,
\end{align}
where the Green's function $\tilde{G}_{ijkl}$ is given by
\begin{align}
    \tilde{G}_{ijkl}=\frac{1}{\omega^2-|\bmath{k}|^2}\sum_{I}
    \left(h^{I*}_{ij} h^I_{kl}+\mbox{(c.c)}\right)\,,
\end{align}
and the quantities associated with $\tilde{}~$ represent Fourier components, 
defined by
\begin{equation}
    \tilde h_{\mu\nu}(k)=\frac{1}{(2\pi)^4}\int d^4\! x\, e^{-ik_\mu x^\mu}h_{\mu\nu}(x)\,.
\end{equation}
Thus, the metric perturbation would be given by
\begin{align}
    h_{ij}(x)& = \int d^4\! k\, e^{ik_\mu x^\mu} \tilde G_{ijkl}(k) \tilde T^{kl}(k)\cr
             & =\sum_I \int d^4\! k\, e^{ik_\mu x^\mu} \frac{h^{I*}_{ij} h^I_{kl}+\mbox{(c.c.)}}{\omega^2-|\bmath{k}|^2}\tilde T^{kl}(k)\,.
\label{eq:metricperturbation}
\end{align}
%The symmetry $\tilde{G}_{ijkl}=\tilde{G}_{jikl},\ \tilde{G}_{ijkl}=\tilde{G}_{ijlk},\ \tilde{G}_{ijkl}=\tilde{G}_{klij}$
%allows all the possible combinations of the three basis tensors $e^{(A}_{ij}e^{B)}_{ij}$ and its dimension becomes $10$. However, 

To respect the rotational symmetry, % for $\tilde{G}_{ijkl}$, 
$\tilde{G}_{ijkl}$ should be composed of the invariant tensors $\delta_{ij}$ and $k^{i}$.  
Imposing the symmetries $\tilde{G}_{ijkl}=\tilde{G}_{jikl},\ \tilde{G}_{ijkl}=\tilde{G}_{ijlk},\ \tilde{G}_{ijkl}=\tilde{G}_{klij}$, the form of $\tilde{G}_{ijkl}$ is restricted to
\begin{align}
 \tilde G_{ijkl}(k)= \frac{1}{(\omega^2-|\bmath{k}|^2)}&
    \Biggl[ a_1 \delta_{ij} \delta_{kl} +a_2 \delta_{i(k} \delta_{l)j}
    +a_3 \hat k_{(i}\delta_{j)(k}\hat k_{l)}\cr
    & +a_4\left(\hat k_i\hat k_j\delta_{kl}+\delta_{ij}\hat k_k \hat k_l\right)
     +a_5 \hat k_i \hat k_j \hat k_k \hat k_l\Biggr]\,,
     \label{eq:Greenfn}
\end{align}
where $\hat k_i=k_i/|\bmath{k}|$ is the unit vector pointing at the wave propagation direction.

The Green's function obtained by summing over $+$ and $\times$ modes, $\tilde G^{(T)}_{ijkl}\propto e^{+}_{ij}e^{+}_{kl}+e^{\times}_{ij}e^{\times}_{kl}$ should satisfy
$k^i \tilde G^{(T)}_{ijkl}=0$ and $\delta^{ij} \tilde G^{(T)}_{ijkl}=0$. 
Combined with the condition 
$\delta^{ik}\delta^{jl}\tilde G^{T}_{ijkl}=2/(\omega^2-|\bmath{k}|^2)$, 
these conditions completely determine the coefficients as 
\begin{equation}
(a^T_1,a^T_2,a^T_3,a^T_4,a^T_5)=(-1,2,-4,1,1)\,.
 \label{eq:tensorcoeff}
\end{equation}
The contribution to the Green's function due to the propagating mode with the scalar modes $h^{3}_{ij}$ would be given 
by 
\footnote{In the same manner, we can also calculate the Green's function due to the propagating mode with the vector modes such as ${h}^{I}_{ij}=\zeta_{x}e^{x}_{ij}+\zeta_{y}e^{y}_{ij}$,
\begin{align}
 (a^V_1,a^V_2,a^V_3,a^V_4,a^V_5)= 
  \left(0, 0, 2\left(|{\zeta}_{x}|^2+|{\zeta}_{y}|^2\right), 0, -2\left(|{\zeta}_{x}|^2+|{\zeta}_{y}|^2\right) \right)\,.
\end{align}
%However, we will focus on the scalar polarization in this paper from the reasons that we mentioned in Sec.~\ref{sec:introduction}.
} 
\begin{equation}
 (a^S_1,a^S_2,a^S_3,a^S_4,a^S_5)= 
  \left(|{\zeta}_b|^2, 0,0,\Re\left[\left(\sqrt{2}{\zeta}_l-{\zeta}_b\right){\zeta}^*_b\right],\left\vert\sqrt{2}{\zeta}_l-{\zeta}_b\right\vert^2\right)\,.
\end{equation}
In the case with multiple massless degrees of freedom, the above formula can be simply extended to the sum of contributions from all modes.

Now, we shall consider the constraint on the polarization modes from the tests of gravity in the solar system. For this purpose, we observe how the linear perturbation around a static spherically symmetric object is reproduced. The above Green's function generates only $ij$-components of the metric perturbation, which are given in a different gauge from the standard Newtonian or post-Newtonian gauge. Hence, the gauge transformation to a more familiar form of the metric is also addressed in the following discussion. 

We start with the $a_5$ term, which turns out to give the dominant component of the Newtonian potential. 
Using the conservation law, we can replace the spatial component of the stress-energy tensor $T^{ij}$ with $T_{00}$ as $k_i k_j \tilde T^{ij}=\omega^2 \tilde T^{00}$. 
Namely, we can calculate the contribution from the $a_5$ term as
\begin{align}
    a_5\int d\omega\, d^3\! k& \frac{k_i k_j k_l k_m}{|\bmath{k}|^4 (\omega^2 -|\bmath{k}|^2)} \tilde T^{lm} e^{ik_\mu x^\mu}\cr
    &=- a_5\partial_i\partial_j\int_{-\infty}^{\infty} d\omega\int d^3\! k \frac{\omega^2\tilde T^{00}}{|\bmath{k}|^4 (\omega^2 -|\bmath{k}|^2)} e^{-i\omega t+ i\sbmath{k} \cdot \sbmath{x}}\cr
    &=-\frac{a_5}{2}\partial_i\partial_j\frac{(2\pi)^2}{r}\int_{-\infty}^{\infty} d\omega \, \omega^{-2}e^{-i\omega(t-r)}\tilde{T}_{00}\cr
    &=a_5\partial_i\partial_j  \frac{M t^2}{8\pi r}\,,    
\end{align}
where in the third equality we replaced $\omega^{-1}$ with $-i\int dt$ and 
substituted $\tilde T^{00}=M\delta(\omega)/(2\pi)^3$, which is  
obtained from $T^{00}=M\delta^3(\mbox{\boldmath{$x$}})$. 
%we define the enclosed total mass $M=\int d^3x \tilde T^{00}$.

The secular growth proportional to $t^2$, which originates from the factor $\omega^{-2}$ in the third line, can be eliminated by a gauge transformation.
Under the infinitesimal coordinate transformation $x^\mu\to x^\mu+\xi^\mu$, 
the metric perturbation transforms as $h_{\mu\nu}\to h_{\mu\nu}-2\xi_{(\mu,\nu)}$. 
By choosing $\xi_\mu$ as 
\begin{equation}
    \xi_0= -a_5\frac{t M}{8\pi r}\,,\qquad \xi_j=a_5\frac{t^2 M}{16\pi}\partial_j\frac1r\,,
\end{equation}
the contribution of the $a_5$ term to the metric perturbation becomes 
\begin{equation}
    h^{(5)}_{00}=-\frac{a_5 M}{4\pi r}\,,\qquad h^{(5)}_{0i}=0\,,\quad h^{(5)}_{ij}=0\,.
\end{equation}
In the same way, we can compute the contribution of the $a_4$ term as 
\begin{align}
    a_4\int d\omega\, d^3\! k& \frac{k_i k_j \delta_{lm}+\delta_{ij} k_l k_m}{|\bmath{k}|^2 (\omega^2 -|\bmath{k}|^2)} \tilde T^{lm} e^{ik_\mu x^\mu}\cr
    &=-a_4\partial_i\partial_j  \frac{\delta_{kl} q^{kl} t^2}{4\pi r}-a_4\delta_{ij}\frac{M}{2\pi r}\,,
\label{eq:contribution_of_a4}
\end{align}
where we define $q^{ij}:=\int d^3x\, T^{ij}$. In the following, we neglect the contributions depending on $q^{ij}$ because $q^{ij}$ can be rewritten as 
\begin{align}
  q^{ij} =\int d^3 x\, \frac{\partial x^i}{\partial x^k} 
    \frac{\partial x^j}{\partial x^l} T^{kl}
     =\int d^3 x\, x^i x^j \partial_t^2 T^{00} + \mbox{(boundary terms)}\,.
\end{align}
Hence, only the boundary terms remain for a stationary source. 
The boundary terms are quadratic in the metric perturbation and 
hence suppressed by the smallness of the Newtonian potential. 
Therefore, we have $q^{ij}=O(\Phi)M$. 
Determination of the perturbation at this order is beyond the scope of the linear theory discussed here. 
Since the contributions of the $a_1$, $a_2$, and $a_3$ terms are proportional to $q_{ij}$ or vanish, we do not discuss these contributions.
As a result, we obtain
\begin{equation}
    h_{00}=-\frac{a_5 M}{4\pi r}\,,\qquad h_{0i}=0\,,\qquad 
    h_{ij}=-\frac{a_4 M}{4\pi r}\delta_{ij}\,.
\end{equation}
Substituting the coefficients of the contributions from the tensor modes \eqref{eq:tensorcoeff}, we can recover the standard result in GR,
\begin{equation}
     h^{(T)}_{00}=\frac{M}{8\pi r}\,,\qquad h^{(T)}_{0i}=0\,,\qquad 
    h^{(T)}_{ij}=\frac{M}{8\pi r}\delta_{ij}\,.
\end{equation}
The deviation from GR in the relative magnitude between the scalar potentials $h_{00}=2\Phi$ and $h_{ij}=2\Psi\delta_{ij}$ is
denoted by $\gamma$~\cite{Will2005}, which is defined by 
\begin{align}
\frac{\gamma-1}{2}:=\frac{\Psi}{\Phi}-1\,,
\end{align}
in the context of the parameterized post-Newtonian(PPN) formalism. $\gamma=1$ corresponds to the GR case. In the present case we have
\begin{align}
    \frac{\gamma-1}{2}&=\frac{a_5}{a_4}-1\cr
       &\approx -2\Re\left[\left({\zeta}_b-\frac1{\sqrt{2}}{\zeta}_l\right)\left({\zeta}^*_b-\sqrt{2}{\zeta}^*_l\right)\right]\,.
       \label{eq:gammahiku1}
\end{align}
Since the additional polarization modes couple to the matter, the new force mediated by the additional modes, which is often called the fifth force, would be present. The strength of the fifth force is constrained by various experiments~\cite{Murata2015}.
The fifth force mediated by the massless field is most strongly constrained by the measurement of the Shapiro time delay by the Cassini satellites~\cite{Bertotti2003}.
The constraint is given by
\begin{align}\label{eq:fifthconst}
	\frac{\gamma-1}{2}<2.1\times 10^{-5}~.
\end{align}
After translating the bound~\eqref{eq:fifthconst} to the bound on the coupling parameter from Eq.~\eqref{eq:gammahiku1} and combining it with the bounds on the energy flux~\eqref{eq:constraintB}, we can constrain the detector amplitude of the signal of additional polarization modes, as we shall see below. 

Since $\psi_I$ 
takes the same standard form as given for $\phi_I$ in \eqref{eq:standardform}, 
$\psi_I$ must be related to $\phi_I$ by a transformation 
\begin{equation}
    \psi_I={\cal U}^J_{~I}\phi_J\,, 
\end{equation}
with a unitary matrix ${\cal U}$. 
Suppose that $\phi_1$ and $\phi_2$ are the results of propagation of the standard two tensor modes evaluated at the position of the observer. 
An additional mode $\phi_3$ may or may not be excited at the source and propagate in the same way as the other two modes. 

When additional propagating mode $\phi_3$ is excited, the amplitude $|\phi_3|$ relative to the two tensor modes $|\phi_1|$ and $|\phi_2|$ is directly 
given by the fractional deviation of the energy loss in Eq.~\eqref{eq:Bq} as
\begin{align}
\frac{|\phi_3|^2}{|\phi_1|^2+|\phi_2|^2}\lesssim B_q v^{2q}\,.
\label{eq:energy_flux}
\end{align} 
On the other hand, as we do not have a generic argument that constrains the form of $U^J_{~I}$, 
in principle, $U^J_{~I}$ can be an arbitrary unitary matrix. 
Hence, Eq.~\eqref{eq:energy_flux} does not constrain the relative 
magnitude of the extra mode when expanded in terms of the observer's basis, 
$|\psi_3|/\sqrt{|\psi_1|^2+|\psi_2|^2}$. 
However, it is impossible to choose the unitary matrix $U$ such that 
$|\psi_3|$ becomes much larger than $\sqrt{|\psi_1|^2+|\psi_2|^2}$, when
the power is almost equally distributed among two or three modes. 
If we have two independent scalar modes, two tensor amplitudes might be 
completely transferred to the two scalar modes. 
Even in that case, we should notice that the interchange 
between tensor modes and scalar modes can occur only under the influence 
of anisotropic background. As long as we consider a homogeneous isotropic background, 
there can be no mixing between different helicity modes. 
Therefore, the maximum amplitude of the energy flux of the scalar mode that we can expect would be the one estimated under the assumption of equi-partition 
among all propagating modes. 
Thus, we conclude that the amplitude of the scalar polarization mode $|\psi_3|$
is, at most, the same order with the tensor modes $\sim \sqrt{|\psi_1|^2+|\psi_2|^2}$,
{\it i.e.}, 
\begin{align}
\frac{|\psi_3|^2}{|\psi_1|^2+|\psi_2|^2}\lesssim 1.
\label{eq:energy_flux2}
\end{align}

%determined by the constraint on the coupling.

Then, the amplitude of scalar polarization measured by GW detectors would be suppressed by the factor $\zeta_b-\sqrt{2}\zeta_l$ from Eq.~\eqref{eq:observerbasis} compared with the tensor polarization. 
This factor is constrained by the constraint on the PPN parameter $\gamma$, Eq.~\eqref{eq:fifthconst}, through Eq.~\eqref{eq:gammahiku1} 
unless the parameters $\zeta_b$ and $\zeta_l$ are tuned. 
Therefore, the upper bound on the detector amplitude of the scalar polarization relative to the tensor polarization 
can be estimated as $O(10^{-2.5})$ from the constraint on the coupling factor with the constraint on the energy flux of Eq.~\eqref{eq:energy_flux2}. 

On the other hand, this constraint could be evaded if the parameters are tuned so as to vanish the right hand side of Eq.~\eqref{eq:gammahiku1}.
This is possible if $|{\zeta}_l|$ is in the range between $\sqrt{2} |{\zeta}_b|$ and $|{\zeta}_b|/\sqrt{2}$ and its phase is chosen appropriately.
%If $|{\zeta}_l|$ is in the range between $\sqrt{2} |{\zeta}_b|$ and $|{\zeta}_b|/\sqrt{2}$, there is a fine-tuned choice of the phase of ${\zeta}_l$ such that the right hand side of Eq.~\eqref{eq:gammahiku1} vanishes. 
The possible choice is  ${\zeta}_b=\sqrt{2}{\zeta}_l$ or ${\zeta}_l=\sqrt{2}{\zeta}_b$. Then, the value of $\gamma-1$ would be  dominated by the contribution coming from $q^{ij}$, which becomes  
$O({\zeta}_b^2 \Phi)$, and can be sufficiently suppressed even if $|{\zeta}_b|$ and $|{\zeta}_l|$ are $O(1)$. In the case with ${\zeta}_b=\sqrt{2}{\zeta}_l$, the sensitivity of the GW detectors to 
the scalar polarization vanishes. Hence, what we are interested in here 
would be the cases close to the opposite boundary with ${\zeta}_l=\sqrt{2}{\zeta}_b$. The fine tuning condition becomes more complicated in the case of multiple massless scalar degrees of freedom. 

Even in such fine-tuned models, in general, the second order correction to the Newtonian potential $\Phi$ will 
have $O({\zeta}_b^2 \Phi^2)$ deviation from the prediction of GR. 
The coefficient of the deviation proportional to $\Phi^2$ is often denoted by $\beta-1$~\cite{Will2005}. 
In terms of $\beta$, $h_{00}$ to the second order is expressed as $2\Phi-2\beta\Phi^2$, and 
$\beta=1$ corresponds to the GR case. 
This $\beta$ parameter is slightly less tightly constrained compared with $\gamma-1$, from the 
measurement of the perihelion advance as $|\beta-1|<8\times 10^{-5}$~\cite{Will2005}. 
The constraint on the detector amplitude is 
relaxed to the one set by the observation of $\beta-1$, which is $O(10^{-2})$, 
although we do not have any concrete model that can realize such a tuning.

\subsection{Detectability of apparent scalar mode induced by background anisotropies}
In the previous subsection, we give an upper bound on the detector amplitude 
of the scalar polarization 
when the observational environment in the solar system can be approximated by a homogeneous and isotropic background.
In this subsection, we consider the possibility of detecting apparent scalar polarization induced by the hypothetical background anisotropies at the stage of the observation. 
Here, we assume the background anisotropies in the solar system, 
but we do not assume any additional polarization modes propagating from the GW source. 

Hence, we start with only two tensor propagating polarization modes at hand. 
We consider that the polarization basis tensors of the two modes would be perturbatively modified by the anisotropic background from the original ones, $e^{(+)}_{ij}$ and $e^{(\times)}_{ij}$. 
The unperturbed $+$ and $\times$ polarization basis tensors have an ambiguity corresponding to  
the degree of freedom for the rotation around the axis pointing $k^i$. 
The final result must be independent of the way how to fix this arbitrariness. 
In order to maintain this basic symmetry, 
the modified polarization tensors should be generated by applying the same operator to the original polarization tensors, $e^{(+)}_{ij}$ and $e^{(\times)}_{ij}$. 
Note that the Green's function 
constructed from the modified polarization tensors violates the rotation symmetry reflecting the background anisotropies.

To describe the homogeneous but anisotropic background, we introduce 
a background homogeneous vector $W^i$, or tensor $X^{ij}$. 
We restrict our attention to the perturbation linear in $W^i$ or $X^{ij}$. 
Then, we can use $\hat k^i$, $k$ and $\delta_{ij}$ as well as $W^i$ and $X^{ij}$ to generate the modified polarization tensors.
Here, we claim that we need to associate $k^0$, $k$ or $k^2$ to $W^i$ and $X^{ij}$, according to the number of indices contracted with $\hat k^i$. When the index of $\hat k^i$ is not contracted with $W^i$ or $X^{ij}$, $k^i$ 
in $\hat k^i$ can be converted to $\omega$ in the end after the manipulations described in the preceding subsection.
Then, $\hat k^i$ does not produce any extra negative power of $\omega$. 
However, once $\hat k^i$ is contracted with $W^i$ or $X^{ij}$, $k^i$ in $\hat k^i$ is replaced with the spatial derivative and 
the factor $1/k$, which produces a negative power of $\omega$.
The extra negative power of $\omega$ results in the secular growth of metric perturbation in time that is not eliminated by the gauge transformation. 
Such an extra negative power of $\omega$ can be avoided simply by associating appropriate power of $k$, whose order is determined by the number of indices in $W^i$ and $X^{ij}$ contracted with $\hat k^i$.
We can multiply higher powers of $k$ more than the minimal requirement to avoid the secular growth. 
However, extra $k^2$ would be transformed to the Laplacian operator, giving vanishing contribution when it acts on $1/r$ or its derivatives.
Thus, we find that the terms that can appear at the linear order in $W^i$ and $X^{ij}$ generated from the tensor modes are given by
\begin{align}
  &\mbox{(a)}~  k W^k e^{(T)}_{k(i} \hat k_{j)}\,,\quad
              k W^k \hat k_k e^{(T)}_{ij}\,,\quad
  \mbox{(b)}~  k^2 X^{kl} e^{(T)}_{kl} \hat k_{i}\hat k_{j}\,,\quad
  k^2 X^{kl} \hat k_{k}\hat k_{l} e^{(T)}_{ij}\,,\quad k^2\hat{k}_{l}X^{kl} e^{(T)}_{k(i}\hat{k}_{j)}\,,\cr
  &\mbox{(c)}~  k^2 X^{kl} e^{(T)}_{kl} \delta_{ij}\,,\quad
  \mbox{(d)}~  k X_{~(i}^{k} e^{(T)}_{j)k} \,, \quad
  \mbox{(e)}~  X^{k}_{~k} e^{(T)}_{ij}\,.
\end{align}
where $e^{(T)}_{ij}$ represents the pair of basis tensors, $e^{(+)}_{ij}$ and $e^{(\times)}_{ij}$. 

The contribution of each term can be identified by noticing that
the products of these two original basis tensors after the summation over + and $\times$ modes 
are given by the form in the square brackets in Eq.~\eqref{eq:Greenfn} with 
the coefficients.~\eqref{eq:tensorcoeff}.
Thus, we find that the (a) terms contribute to the Newtonian potential in the form of 
\begin{equation}
\propto W^k \partial_k\frac{M}{r}\,.   
\end{equation}
The correction to the other components of the metric perturbation is higher order in the slow motion dynamics.
The contribution of the (a) terms 
can be understood as an anomalous dipole moment of the gravitational source.
The dipole perturbation in the direction specified by the constant vector $W^i$, caused by the (a) term,  
can be interpreted as a systematic shift of the position of the center of mass. 
Here, we do not discuss this term in details because it does not produce any contamination to the scalar polarization detected by GW detectors. 
In fact, one can easily find that 
\begin{equation}
   e^{(b)ij} \times k W^k e^{(T)}_{k(i} \hat k_{j)}= 0\,,\qquad
   e^{(l)ij} \times k W^k e^{(T)}_{k(i} \hat k_{j)}= 0\,.
\end{equation}

The (b) terms contribute to the Newtonian potential in the form of 
\begin{equation}
    \propto  X^{kl} \partial_k\partial_l \frac{M}{r}\,.
    \label{eq:potential_for_b}
\end{equation}
This contribution can be interpreted as an anomalous quadrupole moment of the gravitational source. 
Since this effect should cause the correction to the orbital motion of planets and satellites, it is observationally constrained by experiments in the solar system.  
We estimate the bound on the magnitude of the extra contribution to the Newtonian potential Eq.~\eqref{eq:potential_for_b}, based on the experimental results by the GRACE satellites~\cite{Goswami2018}, which are monitoring the Earth's gravitational potential. 
The length scale of the orbit, $R_{\oplus}\sim 10^{9}$cm, is comparable to the wavelength of GWs detected by the ground-based GW detectors. 
The modification to the relative velocity between the two satellites can be estimated by the deviation from the Newtonian potential in GR given as 
$\sim |X|r_{\rm sep}/R^{3}_{\oplus}$, where $|X|$ is the magnitude of 
$X^{kl}$ and $r_{\rm sep}\sim 2.2\times 10^7\unit{cm}$ is the separation between two GRACE satellites.  
The observed power spectrum of the relative velocity error 
$\lesssim 10^{-10}$ 
%at the orbital period $90$min 
gives a rough estimate for the constraint on the magnitude, 
$|X|\lesssim 10^{-10}R^{3}_{\oplus}/r_{\rm sep}$. 
Then, we can estimate the bound on the relative 
amplitude of the induced scalar mode as
\begin{align}
\frac{|h_s|}{|h_T|}\sim |X|\omega^2< 10^{-10}\times \frac{R^{3}_{\oplus}\omega^2}{r_{\rm sep}}
%\sim \frac{10^{-7}\ {\rm m/s}}{10^3\ {\rm m/s}}\left(\frac{200\ {\rm Hz}}{50\ {\rm Hz}}\right)^2
\sim10^{-7}\,,
\end{align}
where $|h_s|$ and $|h_T|$ are, respectively, the strain amplitudes of 
the scalar and tensor modes, and 
we set the GW frequency $\omega$ to the typical value, $200{\rm Hz}$, 
corresponding to the observation band of the ground-based GW detectors. 

The (c) term does not provide the correction to the Newtonian potential. Hence, it will not be severely constrained by the solar system gravity test, 
but the (c) term does not give the detectable scalar modes, either. 
This is because the waves induced by the (c) term are proportional to $\delta^{ij}$, while the detector tensor for the interferometric GW detectors is traceless.

The contribution of the (d) term to the gravitational field is very suppressed by the slow motion parameter, since it appears only in the $0i$ component of the metric perturbation. Thus, the observational constraint would be weaker by the factor $\beta\sim10^{-4}$, which is the typical value of the velocity in the solar system. 
The waves induced by the (d) term contains the breathing mode, as can be seen from 
\begin{equation}
    e^{(b)ij} \times k X_{~(i}^{k} e^{(T)}_{j)k}= k X^{kl} e^{(T)}_{kl}\ne 0\,,
\end{equation}
but no longitudinal mode. 
We perform an order estimate similar to the case of the (b) term. 
The magnitude of the metric perturbation $h_{0i}$ relative 
to the Newtonian potential is 
$|X|/R_{\oplus}$. The effect of this perturbation on the 
satellite motion is suppressed by $\beta$. 
Hence, the modification to the relative velocity between the two satellites is estimated as $\sim \beta |X|r_{\rm sep}/R_{\oplus}^2$. 
Then, we can estimate the bound on the relative amplitude of the induced breathing mode as
\begin{align}
\frac{|h_s|}{|h_T|}\sim |X|\omega< 10^{-10}\times \frac{R^{2}_{\oplus}\omega}{r_{\rm sep}}
%\sim \frac{10^{-7}\ {\rm m/s}}{10^3\ {\rm m/s}}\left(\frac{200\ {\rm Hz}}{50\ {\rm Hz}}\right)^2
\sim10^{-3}\,. 
\end{align}
Since the (e) term is proportional to the original tensor basis, the discussion does not change essentially from the case of isotropic background. Trivially, the term does not possess any scalar modes.

To conclude, extra scalar GW polarization modes induced by the hypothetical local anisotropic background would hardly be detected in all of these cases. 

\section{Discussions and conclusion}
\label{sec:Discussions-Conclusion}
We discussed the detectability of scalar polarization modes in GWs from compact binary coalescences. 
We consistently considered the whole processes: 
generation, propagation and observation, in a general framework 
that includes all polarization modes, 
without relying on specific models. 

First, we claimed that the energy flux that can be attributed to additional modes is, at most, comparable to that of the ordinary 
tensor modes, so as to be consistent with the observed GW phase evolution.
Next, we constructed a linear model of GW propagation with full mixing among various polarization modes. 
We showed that the amplitude of each polarization mode may change during propagation, but the energy flux of the propagating modes cannot change without frequency-dependent waveform deformations
as far as the background is smoothly varying and the propagation speed is very close to the speed of light. 
If the dispersion relation is modified significantly, we cannot use the GR waveform as the search templates and need to construct a waveform model considering the propagation process.
If we allow these conditions to be violated, there might be some unknown conversion mechanism, which may cause the enhancement of 
the energy fluxes of the propagating modes. 
Practically, the only possibility would be 
violating the smoothness of the background.  
%the mixing between the different helicity polarization modes requires 
%the background must be not only anisotropic but also inhomogeneous. 
%of the background. 
Anyway, it seems very challenging to find a mechanism that realizes the selective amplification of the energy flux of some particular mode 
without modifying the dispersion relation. 

In our general framework, the mode conversion is not prohibited. 
Hence, the maximum of the energy flux of the scalar mode is given 
by the one estimated by assuming equipartition among all polarization modes. Therefore, the maximum energy flux for the scalar modes that we can expect can be as high as the tensor modes, although we did not present any example of such an efficient conversion mechanism. 
Together with the constraint on the energy flux based on the consideration in the generation process described above, the maximum amplitude of the energy flux of scalar modes at the position of observers would be $O(1)$ relative to that of the tensor modes. 

On the other hand, the deviation from the post Newtonian gravity in GR has been tightly constrained by the several experiments in the solar system.
These constraints are translated into the constraints on the coupling between the scalar mode and GW detectors or the anisotropies of the background field in the solar system. 
As for the observation process, we considered two scenarios: (i) additional scalar modes exist but no background anisotropies exist, and (ii) some background anisotropies exist but no scalar modes are generated and propagate.
In the scenario (i), we used the constraint that the 
energy flux of the scalar modes is, at most, of $O(1)$ relative to 
that of the tensor modes, to give an upper-bound on the amplitude of 
the scalar polarization detected by GW detectors. 
%indicate that the possible amplitude of the scalar modes, which has been probed by several polarization search, is determined by the constraints in the observation process by the experiments in the solar system.
We showed that the scalar polarization modes in the GW signal should be smaller by the suppression factor of $O(10^{-2.5})$ than the tensor polarization modes, based on %the constraints on the Newtonian potential by 
the experiments in the solar system.
In the scenario (ii), the suppression factor was estimated to be as small as $O(10^{-3})$. 

According to~\cite{Takeda2018, Takeda2019}, the detection limit to the additional polarization amplitude is roughly given by the inverse of the signal-to-noise ratio. The detection limit for a single compact binary coalescence event with the ground-based GW detectors would be expected as $\lesssim 10^{-2}$.
Therefore, these suppression factors indicate that it is difficult to detect the scalar polarization modes with a single compact binary coalescence event.
The detectability of scalar polarization modes with ground-based GW telescopes in a single event is severely restricted. 
However, the detection limit using an ideal stacking of multiple events is estimated as $ \sim 10^{-3}$~\cite{Takeda2019}, 
assuming observations of expected multiple events with the third-generation GW detectors such as Einstein Telescope~\cite{Punturo2010} and Cosmic Explore~\cite{Abbott2017c}. 
Hence, we might be able to obtain meaningful constraints on the amplitude of scalar polarization in the future. 
For this purpose, it is required to develop some efficient stacking methods.

\section*{Acknowledgements}
We would like to thank Kei-ichiro Kubota, Atsuhi Nishizawa, Mori Shota, Shinji Tsujikawa, and Miguel Zumalacarregui for useful comments. H.T. is supported by JSPS KAKENHI Grant Nos. 21J01383 and 22K14037. Y.M. is supported by the establishment of university fellowships towards the creation of science technology innovation JPMJFS2123, JSPS Overseas Challenge Program for Young Researchers, and the Sasakawa Scientific Research Grant from The Japan Science Society.
H.O. is supported by Grant-in-Aid for JSPS Fellows JP22J14159.
T.T. is supported by JSPS KAKENHI Grant Nos. JP23H00110 and JP20K03928. 

\bibliography{fifth_force}
\bibliographystyle{ptephy}

\end{document}